\documentclass[aps,prb,twocolumn,floatfix]{revtex4}
\usepackage{graphicx}
\usepackage{bm}
\usepackage{lscape,graphicx}
\usepackage[dvips]{epsfig}
\usepackage{amsmath}

\begin{document}
\title{Bichromatically driven double well: parametric
perspective of the strong-field control 
landscape reveals the influence of chaotic states}
\author{Astha Sethi and Srihari Keshavamurthy}
\affiliation{Department of Chemistry, Indian Institute of Technology, 
Kanpur, India 208 016}

\begin{abstract}
The aim of this work is to understand the influence of
chaotic states in control problems involving strong fields.
Towards this end,
we numerically construct and study the 
strong field control landscape of 
a bichromatically driven double well. 
A novel measure based on
correlating the overlap intensities between Floquet states and
an initial phase space coherent state
with the parametric motion of the quasienergies 
is used to construct and interpret the landscape features.
``Walls" of no control,
robust under variations of the relative phase between the fields,   
are seen on the control landscape and 
associated with multilevel interactions involving chaotic Floquet states. 
\end{abstract}

\maketitle

\section{Introduction}
Controlling quantum phenomena in atomic and molecular systems is
an active area of research today. 
Persistent theoretical and experimental
efforts have led to the formulation of several ingenious control
schemes\cite{ts98,bsbook,zrbook,rz00} for both atomic and molecular systems.
Recently researchers have been interested in determining
the ``control landscape" for specific systems
\cite{rhr04,cr07,dbm96,ggbm96,sgsj04,gm88,grafe07,wells05,wolle05}.
The motivation for
undertaking such extensive studies is atleast twofold. Firstly, several
optimal control efforts\cite{rhr04,cr07,ymr03} 
seem to succeed despite employing
local search algorithms in a prohibitively high dimensional control 
parameter space. 
Rabitz and coworkers\cite{rhr04,cr07}
reason that the topology
of the control landscape
can shed light on the robustness
of the control strategies. 
Secondly, many control schemes\cite{zrbook}
involve few, carefully chosen, levels which
provide valuable insights into the control process.
However, such few-level schemes might be compromised due to several
reasons\cite{cdb07,han98,gb05,lgw94,nr05} and one hopes that
identifying the features of the control landscape associated
with the deviations might lead to better control techniques.

In order to set the stage for our work we briefly discuss
the concept of a control landscape.
Consider a Hamiltonian of the form $H=H_{0}-\mu {\cal E}(t)$ 
with $\mu$ being the dipole operator and ${\cal E}(t)$ being the
control field. 
There are two perspectives on the control problem
which are of interest to the present work. The first one
involves using a given field 
${\cal E}(t;{\bm \lambda})$ characterized by one or
more control variables ${\bm \lambda} = (\lambda_{1},\lambda_{2},\ldots,
\lambda_{f})$. 
The objective in this case is to study the
variation of, say, the transition 
probability $P_{if} \equiv|\langle f|U(T,0)|i\rangle|^{2}$
between an initial state $|i\rangle$
and a final state $|f\rangle$ at some target time $T$. 
The control landscape is constructed
by viewing the transition probability as a functional of the
field parameters {\it i.e.,} $P_{if}=P_{if}[{\cal E}(t;{\bm \lambda})]$.
Regions on the landscape with high values for $P_{if}$ yield the
best possible choice of the control variables in the presupposed
${\cal E}(t;{\bm \lambda})$. Note that it is quite possible to have
several distinct choices for the control variables which yield the
same value for $P_{if}$. Several such examples of control landscapes
can be found in the literature\cite{bsbook,dbm96,ggbm96,gm88,grafe07,sgsj04}

The second perspective on the control problem comes from the
optimal control theory (OCT) approach\cite{rz00,cr07}.
In OCT a cost (landscape) function of the form
$\Phi(U) =
{\rm Tr}[U \rho(0) U^{\dagger} \Theta] \equiv \langle \Theta \rangle$, 
with $\rho(0)$ being
the initial density matrix and an arbitrary observable $\Theta$, is
extremized with respect to the field ${\cal E}(t)$.
The optimal fields correspond to the solutions of the variational
condition $\delta \Phi(U)/\delta {\cal E}(t) = 0$ wherein the
unitary time evolution operator $U=U[{\cal E}(t)]$ is naturally
considered as a functional of the field.
The transition probability
$P_{if}$ is a special case of $\Phi(U)$ with $\rho = |i\rangle \langle i|$
and $\Theta = |f\rangle \langle f|$.
Again, the set of optimal fields
that yield the same $\langle \Theta \rangle$ constitutes
a control level set 
whose topology is
of considerable interest since one can imagine `morphing' an 
experimentally undesirable optimal field to a more desirable field.
In order to be able to carry out
such a morphing it is essential to have full control of the system. 
Assuming full controllability Rabitz and coworkers have 
explored the topology of the OCT landscapes in great detail\cite{cr07}.
A particularly striking result is that 
the full quantum control landscape is devoid of false traps or suboptimal
extrema\cite{rhr04,cr07}.
However, in practice one is invariably
forced to impose constraints on $\Phi(U)$ in various forms. Examples include
specification of target time $T$, minimizing the field fluence, and
reduced or restricted parametrizations of the control field.

Among the two perspectives on control outlined above we adopt the first
one in this work {\it i.e.,} the nature of the control landscape
with a given form of the field.
Although we do not undertake an OCT
calculation in the present work,
it is nevertheless
useful to think of the field used herein as a result of a reduced
dimensional parametrization of an optimal field
coming out of a OCT calculation.
The reason being that 
restricted parametrizations of the optimal
${\cal E}(t)$ typically projects the full OCT control space to a lower
dimensional space which can exhibit suboptimal features\cite{cr07}.
`Suboptimal' features on the
control landscape are, therefore, expected
in our analysis.
Clearly, in the OCT context, 
suboptimal features can be avoided if the restricted parametrization
is done 
carefully by preserving the overall intrinsic topology
of the control landscape and level sets. Discovering such appropriate
parametrizations might not be easy but a recent combined experimental
and theoretical work
by Roslund {\it et al.} provides a clear example\cite{rrr06}.
In certain instances, as in the work by Bartelt {\it et al.}, a systematic
study of simple parametrizations of the optimal field can provide
mechanistic insights into the control process\cite{bfw05}.

Based on the discussion above, one can interpret the suboptimal
features as artificial due to inappropriate field parametrization
or as genuine due to the unavoidable constraints imposed on the
system. 
Consequently, the model study presented in this work incorporates
two of the constraints that are expected to result in a complicated
control landscape - strong fields and simple forms of the control field.
It has been
suggested that in strong field regimes local traps may appear more
readily. However, even in the strong field regimes recent studies indicate
that interpreting the various features of the constrained field control
landscape is far from being straightforward\cite{grafe07,wells05,wolle05}. 
In this context, it is well known that the classical phase space
of strongly driven systems exhibits widespread chaos.
Could it be that local traps on the strong field control landscapes
arise due to the chaos? Unfortunately, not much is known about the
impact of chaotic or delocalized quantum states on 
the topology of control landscapes and level sets.
Such an analysis for the full OCT landscape
is far from being simple. Therefore, in this work
we make a beginning, within the perspective discussed above,
to understand the nature and origins of the suboptimal features from a
classical-quantum correspondence viewpoint.

In most physical systems the phase space is mixed regular-chaotic and
there is sufficient evidence in the literature to indicate that quantum
states delocalized in the chaotic regions can significantly influence the
control of initial states localized in the regular regions. 
For instance, Na and Reichl\cite{nr05} have shown that adiabatic
passage mechanisms are modified due to the presence
of chaotic states leading to chaos-assisted passage.
Gong and Brumer\cite{gb05}
have recently reviewed the subject of
coherent control of quantum phenomenon in the presence of
chaos. Their studies, especially using the (modified) kicked rotor,
have highlighted several intriguing features like faster-than-classical
quantum anomalous diffusion and the nontrivial influence of classical
phase space structures with area much smaller 
than the effective Planck constant.
As a last example we mention the work by Takami and Fujisaki\cite{taka07}
on optimal control of random matrix systems. Random matrix systems
correspond to strongly chaotic classical phase space with no regular
regions. In such cases it is possible, using OCT, to obtain an analytic
expression for the optimal field which controls the transition between
two random states.
The aforementioned studies, and other earlier investigations, concerned
with the control of quantum processes in classically chaotic
systems have raised many interesting issues.
However, there are very few attempts to analyze the various features
on the control landscape in terms of the phase space nature
of the quantum states. Controllability of quantum dynamics
despite the strong chaos is surprising and
calls for a careful study of the control landscape from a classical-quantum
correspondence perspective.

We begin in section~\ref{model} with a brief description of the model
system, driven double well, which has served as
a paradigm for many coherent control studies.
The controllability of the dynamics using a bichromatic field is studied 
as a function of field parameters.
The role of chaotic (delocalized) Floquet states in attempting to
control the decay of an initial state localized in the regular
region of the phase space is highlighted.
Section~\ref{corr} introduces a measure
which involves correlating 
the response of the Floquet states 
to the control field with the
overlap intensities between the initial state and the Floquet states.
We show that the correlator is related to generic landscape functionals
and can therefore yield useful insights
into the features of the control landscape from a classical-quantum
correspondence point of view.
In section~\ref{results}  
we show that the proposed correlator is very sensitive to the
underlying phase space structure and clearly identifies regions
of no control involving delocalized (chaotic) Floquet states.
Section~\ref{summary} concludes with comments on the role
of the chaotic states in the process of control and future studies.

\section{Model system}
\label{model}

The model of choice to illustrate our approach is the driven double
well system. The Hamiltonian is given by
\begin{equation}
H(x,p;t)=\frac{1}{2m}p^{2}+Bx^{4}-Dx^{2}+x{\cal E}(t) \label{bichromham} 
\end{equation}
with ${\cal E}(t)$ being the field. The driven double well model
is an ideal testbed
for several schemes that have been suggested for controlling coherent
superposition of quantum
states\cite{gh98,bsbook,gdjh91,Holth92,fm93,sgsj04}.
Apart from being used to elucidate
phenomena like chaos-assisted tunneling\cite{lb90} and coherent destruction
of tunneling\cite{gdjh91}, the model above is also relevant
in research areas like quantum computing\cite{qc}
and quantum dots\cite{tkp07}.
In this context we also note an early study\cite{sr93} by Schwieters and Rabitz
on the above system with the aim of designing an optimal ${\cal E}(t)$
to reduce the difference between quantum observables and their
classical analogs. In the present work, as alluded to in the introduction,
our goal is to study control using a presupposed field and the
resulting influence on the control landscape. A variety of choices
can be made for ${\cal E}(t)$ and in this work we choose a bichromatic
form 
\begin{equation}
{\cal E}(t)=\lambda_{1} \cos(\omega t) 
+ \lambda_{2} \cos(2\omega t+\phi) \nonumber
\end{equation}
The above choice for ${\cal E}(t)$ is motivated by a large number  
of studies\cite{hcu07} that have shown the utility of such fields 
in controlling the dynamics with
the possibility of using the relative phase $\phi$ as an extra
control knob.
Moreover, due to the periodicity of Eq.~\ref{bichromham},
$H(x,p,t)=H(x,p,t+T_{f})$, with period $T_{f}=2\pi/\omega$,
the dynamics can be conveniently analyzed\cite{pg92} using
Floquet theory\cite{sh65,sa73} in terms of
the Floquet states $\{|\chi_{n} \rangle\}$ and the associated
quasienergies $\{E_{n}\}$. 

\begin{figure} [hbtp]
\includegraphics[height=80mm,width=90mm]{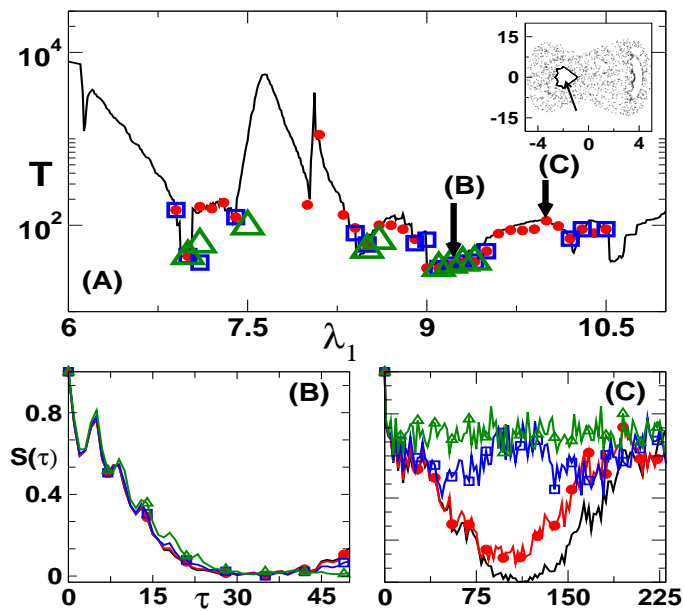}
\caption{(Color online) (A) Decay time $T$ versus
the field strength $\lambda_{1} (\lambda_{2}=0)$ for the coherent state,
$|z\rangle \equiv |x_{0},p_{0}\rangle = |-1.5,0\rangle$,
localized in the left regular island (indicated by an arrow in
the inset showing the stroboscopic phase space at $\lambda_{1}=6.4$).
(B) and (C) show the influence of the $2\omega$-control field
with $\phi=0$ on the survival
probabilities for $\lambda_{1}=9.2$ and $10$
respectively. Increasing $\lambda_{2}$ from $0$ (black)
to $0.2$ (red circles), $0.7$ (blue squares), and $2.1$ (green triangles)
leads to localization in case (C) but not in case (B). The same information,
with consistent color code,
is shown in (A) as points unaffected by increasing $\lambda_{2}$.}
\label{fig1}
\end{figure}

Throughout this work $\hbar=1$ and we fix the
parameter values\cite{lb90,fm93}
$m=1, D=10, B=0.5$ and, $\omega=6.07$.
The double well supports about eight tunneling doublets
and $\omega$ nearly corresponds to the energy separation
between the ground and first excited states, neglecting
the tunnel splittings.
The primary field strength $\lambda_{1} \in [6.0,11.0]$ is chosen such that
the classical phase space, for $\lambda_{2}=0$,
has two symmetry related regular islands
embedded in a chaotic sea (cf. inset in Fig.~\ref{fig1}A).
The symmetry related regular regions and the associated symmetric pairs
of Floquet states exist due to the
discrete symmetry
$H(p,x,t) = H(-p,-x,t+\pi/\omega)$.
The parameter values have been chosen to make contact
with an earlier work by Farrelly and Milligan\cite{fm93}
which explored the possibility of controlling the tunneling
using bichromatic fields.
Different choices for
the parameters, as long as the phase space exhibits substantial chaos,
show similar behaviour and do not alter our conclusions.

\subsection{Control landscape: influence of the chaotic states}
\label{clandscape}

\begin{figure} [t]
\includegraphics[height=80mm,width=80mm]{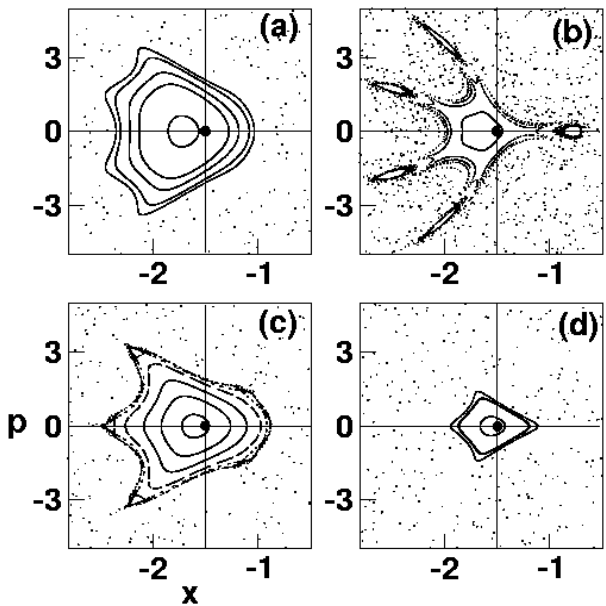}
\caption{The local nature of the phase space in the vicinity of the
initial state $|z\rangle$ (black circle) for $\lambda_{1}$ equal to
(a) 7.0 (b) 8.0 (c) 9.0 and (d) 10.0 with $\lambda_{2}=0$.
Note the drastic change in (b) which is mirrored in the decay time
behavior in Fig.~\ref{fig1}A.}
\label{fig2}
\end{figure}

The control problem that we address is as follows.
The initial state, as in the earlier studies\cite{lb90,fm93},
is a coherent state $|z\rangle \equiv
|x_{0},p_{0}\rangle$ localized in the phase space about 
$(x_{0},p_{0})$
within one of the regular islands. 
The coherent state is constructed as linear superposition
\begin{equation}
|x_{0},p_{0}\rangle = \sum_{b=0}^{\infty} c_{b}(x_{0},p_{0}) |b \rangle
\end{equation}
of a harmonic oscillator basis $\{|b\rangle \}$ with
frequency $\omega_{b}=6.25$ 
(corresponding to $\langle (\Delta x)^{2} \rangle = 0.08$).
The coefficients are given by 
\begin{eqnarray}
c_{b}(x_{0},p_{0}) &=& \frac{1}{\sqrt{b!}} \alpha^{b} e^{-|\alpha|^{2}/2}
e^{-ix_{0}p_{0}/2} \\
\alpha &=& \sqrt{\frac{\omega_{b}}{2}} \left(x_{0} + \frac{i}{\omega_{b}}
p_{0} \right) \nonumber
\end{eqnarray}
We employ the same basis to compute the Floquet states and quasienergies
used in the rest of this work.
The time evolved state $|z(t)\rangle$ can be expressed in terms of the
Floquet states as
\begin{equation}
|z(t)\rangle = \sum_{n} e^{-iE_{n}t} |\chi_{n}(t)\rangle
\langle \chi_{n}(0)|z(0)\rangle
\end{equation}
Measuring time in units of $T_{f}$ and owing to the periodicity
of the Floquet states, $|\chi_{n}(t)\rangle=|\chi_{n}(t+T_{f})\rangle$,
the above equation simplifies to
\begin{equation}
|z(\tau)\rangle = \sum_{n} e^{-iE_{n}\tau} |\chi_{n}(0)\rangle
\langle \chi_{n}(0)|z(0)\rangle
\label{zper}
\end{equation}
with $\tau \equiv k T_{f}$ and integer $k$.
In the absence of the $2\omega$-field
(``control field") $|z\rangle$ decays over a timescale $T$ 
which is determined from  
the first vanishing of the  survival probability
\begin{eqnarray}
S(\tau) &\equiv& |\langle z(0)|z(\tau) \rangle|^{2} \nonumber \\
&=&\left|\sum_{n} e^{-iE_{n}\tau} \langle z(0)|\chi_{n}(0)\rangle
\langle \chi_{n}(0)|z(0)\rangle \right|^{2} \nonumber \\
&=& \sum_{m,n}^{} p_{zn}p_{zm} e^{-i(E_{n}-E_{m})\tau}
\label{survprob}
\end{eqnarray}
where the overlap intensities 
are denoted by $p_{zn} \equiv |\langle z(0)|\chi_{n}(0) \rangle|^{2}$.
Note that $S(\tau)$ is just a special case of $P_{if}$ with
$|i\rangle=|f\rangle=|z\rangle$.
In Fig.~\ref{fig1}A the decay time is shown as a function of
$\lambda_{1}$ for $\lambda_{2}=0$. 
The strong fluctuations in $T$, despite similar phase space
nature over the entire range of $\lambda_{1}$, is a
direct consequence of the underlying chaos\cite{lb90}.
Recently it has been established\cite{es05,sk05} that the local
structure of the phase space about the location of $|z\rangle$
can play a crucial role in the decay dynamics. 
The local phase space structure around the left regular island
is shown in Fig.~\ref{fig2} for various values of $\lambda_{1}$.
Comparing the features of the decay time plot in Fig.~\ref{fig1}A
with the local phase space nature in Fig.~\ref{fig2} shows  
qualitative correspondence.
In particular, the drastic variation of $T$ around $\lambda_{1} = 8$,
seen in Fig.~\ref{fig1}A,
is reflected in the significant change of the local phase space 
nature in Fig.~\ref{fig2}(b) as compared 
to those in Fig.~\ref{fig2}(a),(c), and (d).
There is also preliminary evidence (not discussed in this work)
which indicates that the various field-matter nonlinear
resonances control the decay of $|z\rangle$ and partly responsible
for some of the structure seen in Fig.~\ref{fig1}A.

Is it possible to control the decay of $|z\rangle$ with an appropriate
choice of the parameters $(\lambda_{2},\phi)$ of the $2\omega$-field?
In this instance the
problem of control, in the asymptotic limit $\tau \rightarrow \infty$,
is equivalent to the suppression of tunneling
between the two symmetry related islands in the phase space. 
At the outset one expects, for $\lambda_{2} \neq 0$ and $\phi \neq \pi/2$,
the discrete symmetry to be broken
leading to localization
and hence perfect control.
However, the results of the computations summarized in Fig.~\ref{fig1}
show a more complicated behaviour. 
For most of this study we will choose two representative cases
to illustrate the issues.
In the first case, corresponding to $\lambda_{1} = 10.0$ and
labeled as (C) in Fig.~\ref{fig1},
a monotonic increase of $\lambda_{2}$ with $\phi=0$ results in the
localization of $|z\rangle$. This is in accordance with the earlier
observations by Farrelly and Milligan\cite{fm93}. 
On the other hand,
the decay dynamics for the second case,
corresponding to $\lambda_{1} = 9.2$ and labeled as (B)
in Fig.~\ref{fig1}, is unaffected
by the control field even for fairly large values of $\lambda_{2}$. 

\begin{figure} [hbtp]
\includegraphics[height=80mm,width=90mm]{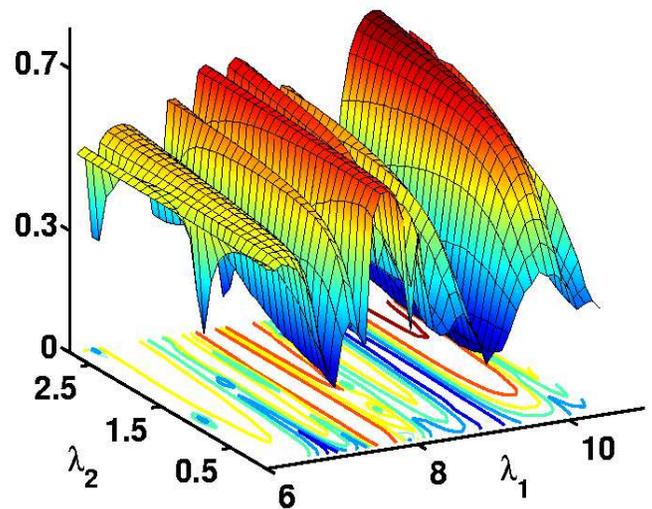}
\caption{(Color online) Time smoothed survival probability
representation of the control landscape for $\phi=0$. The initial state
is the same as in Fig.~\ref{fig1}. Notice the convoluted form of the
landscape with the regions of low probability indicating little to no
control.}
\label{fig3}
\end{figure}

To obtain a global picture,
in Fig.~\ref{fig3} the control landscape for $\phi=0$ is
shown in terms of the time-smoothed survival probability
\begin{equation}
\langle S \rangle = \frac{1}{\tau} \int_{0}^{\tau} d \tau' S(\tau')
\end{equation}
as a function of the field parameters $(\lambda_{1},\lambda_{2})$.
Note that the choice of $\langle S \rangle$ to represent the
landscape is made for convenience; the decay time is a better choice
which requires considerable effort but the gross qualitative features of
the control landscape do not change upon using $\langle S \rangle$.
Large (small) values of $\langle S \rangle$ indicate that the decay dynamics
is suppressed (enhanced). It is clear from Fig.~\ref{fig3} 
that the landscape is
highly convoluted - a consequence of the simple bichromatic choice for
${\cal E}(t)$ and the nonlinear nature of the classical dynamics. 
An earlier OCT study\cite{sr91} on a different
classically chaotic system has also suggested
that the control landscape could be highly convoluted.
From a control point of view, there are regions on the
landscape for which a monotonic increase of $\lambda_{2}$ leads to
increasing control. 
At the same time there are regions,
especially around $\lambda_{1} \approx 9.2$, where increasing
$\lambda_{2}$ hardly effects the decay dynamics and
indicates very little control.
As noted before, one expects such features even on constrained
OCT landscapes. 
In this work we would like to understand the nature and origin of such
robust regions of no control.

\begin{figure} [hbtp]
\includegraphics[height=120mm,width=90mm]{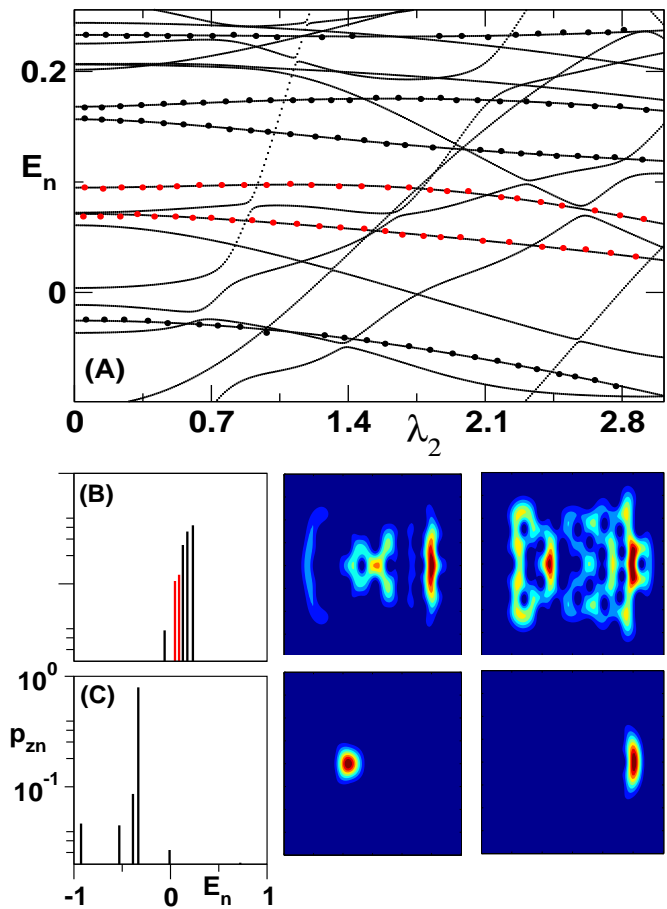}
\caption{(Color online) (A) Variation of the
Floquet quasienergies with
$\lambda_{2}$ and fixed
$\lambda_{1} = 9.2$. Six states that have appreciable
overlap with $|z\rangle$ are highlighted by circles.
(B) Overlap intensity $p_{zn}$ for $\lambda_{1}=9.2$ and
$\lambda_{2}=2.1$ indicates multilevel interactions involving
the states shown in (A). Husimi distribution function of
two delocalized
Floquet states (red) are also shown. (C) Same as in (B)
with $\lambda_{1} = 10$ and $\lambda_{2} = 0.7$ shows
the localized nature of the participating states.}
\label{fig4}
\end{figure}

Insights into the lack of control
can be obtained by studying the
variation of the Floquet quasienergies
with $\lambda_{2}$. 
In Fig.~\ref{fig4}A we show the results of such a computation
corresponding to region B
(cf. Fig.~\ref{fig1}).
Over the entire range of $\lambda_{2}$ six states contribute 
to the decay of $|z\rangle$ as evident from the 
overlap intensities shown in Fig.~\ref{fig4}B.
A closer look reveals the existence of states delocalized
in the chaotic regions of the phase space - two such states are
shown in Fig.~\ref{fig4}B in terms of their Husimi distributions\cite{lb90}.
In contrast, Fig.~\ref{fig4}C shows that the dynamics
of case (C) in Fig.~\ref{fig1} is dictated
by localized states which explains the ease of control with the
symmetry breaking $2\omega$-field. 
Therefore it is reasonable to expect that the lack of control,
signalled by plateaus in Fig.~\ref{fig1}A and the valleys in Fig.~\ref{fig3},
is due to the dominant participation by
chaotic states {\it i.e.,} chaos-assisted tunneling\cite{tu94}.
The plateaus arise due to the fact that the coupling between the 
localized states and the delocalized states
vary very little with increasing control field strength $\lambda_{2}$ -
something that is evident from the Floquet level motions shown in
Fig.~\ref{fig4}A and established earlier by Tomsovic and Ullmo in their
seminal work\cite{tu94} on chaos-assisted tunneling 
in coupled quartic oscillators.
It is important to note that for $\lambda_{2}=0$
the chaotic states, as opposed to the regular states,
do not have a definite parity. Consequently, presence of the
$2\omega$-field does not have 
a major influence on the chaotic states\cite{lgw94}.
Thus, if one or more chaotic states are already influencing the dynamics
of $|z\rangle$ at $\lambda_{2}=0$ then the bichromatic control is
expected to be difficult. 

Similar conclusions were arrived at in a previous
study\cite{lgw94} involving bichromatically driven pendulum.
However, several questions, not addressed by the previous studies, 
arise at this juncture. What is the precise role
of the chaotic states for control? Are the
plateaus robust for varying $\lambda_{2}$ and $\phi$? Is there a
suitable measure, generalizable to systems with higher degrees of freedom,
that can identify regions on the control landscape with little or no control?

\begin{figure*} [hbtp]
\includegraphics[height=90mm,width=180mm]{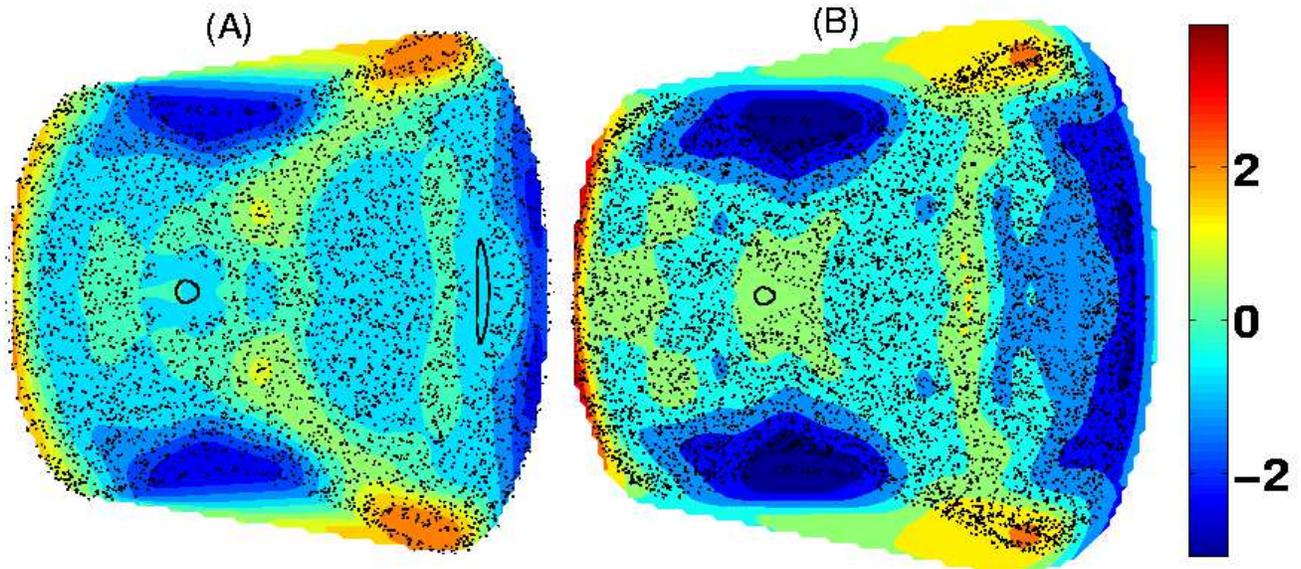}
\caption{(Color online)
The intensity-velocity correlator (cf. Eq.~\ref{intvelcor}) computed over
the entire phase space at $(\lambda_{2}=0.7, \phi=0)$ 
for (A) $\lambda_{1}=9.2$
(B) $\lambda_{1} = 10.0$. Calculations are performed by placing coherent
states at the different phase space points. Note the similarity of
the phase space but the essential differences in the correlator maps.
See text for details.}
\label{fig5}
\end{figure*}

\section{Sensitive measure for the control landscape}
\label{corr}

In order to address the questions raised above we first suggest a
measure, different from $S(\tau)$ or $\langle S \rangle$,
to map out the control landscape. 
Note that the Floquet states and quasienergies of  
Eq.~\ref{bichromham} are parametrically dependent on
the field strengths {\it i.e.,} $E_{n}=E_{n}(\bm{\lambda})$ and
$|\chi_{n}\rangle = |\chi_{n}(\bm{\lambda})\rangle$.  
Thus the response of a Floquet state to the $2\omega$-field, as shown
in Fig.~\ref{fig4}A, is
measured by the parametric derivative 
$\partial E_{n}/\partial \lambda_{2}$ which is also called as the 
level ``velocity". 
At the same time the decay dynamics of a  
coherent state $|z\rangle$ is dominated by
Floquet states that have appreciable 
overlap with $|z\rangle$ (cf. Eq.~\ref{survprob}).
Therefore, Floquet states $|\chi_{n}\rangle$ with substantial overlap
$p_{zn}$ and finite $\partial E_{n}/\partial \lambda_{2}$
are expected to be important in controlling the dynamics of $|z\rangle$. 
The qualitative argument can be made quantitative by introducing the
overlap intensity-level velocity correlator 
\begin{equation}
C_{z}(\lambda_{1},\lambda_{2}) = \frac{1}{\sigma_{z}\sigma_{E}}
\left \langle p_{zn} \frac{\partial E_{n}(\bm{\lambda})}
{\partial \lambda_{2}} \right \rangle_{n}
\label{intvelcor}
\end{equation}
$C_{z}(\bm{\lambda})$ was first proposed by Tomsovic\cite{Toms96}
as a sensitive measure for deviations from ergodicity in strongly
chaotic systems and has been studied in detail in subsequent works\cite{kct02}.
In Eq.~\ref{intvelcor}, $\sigma_{z}^2$ and $\sigma_{E}^2$ are the local
variances of $p_{zn}$ and $\partial E_{n}/\partial \lambda_{2}$ respectively
and the average is
over all the Floquet states. 

The sensitivity of the correlator to phase space structures has been
amply demonstrated in the previous studies\cite{kct02}. Since the earlier
works did not explicitly deal with driven systems, in Fig.~\ref{fig5}
we give one example of the sensitivity in the present context. 
In Fig.~\ref{fig5} we place coherent states at different phase space
points on the phase space and obtain a correlator map. Note that the
coherent state parameters are identical to the ones used in the earlier
sections. The correlator map is shown for two different field strengths
(A) $(\lambda_{1},\lambda_{2})=(9.2,0.7)$ and (B) $(10.0,0.7)$
with $\phi=0$. We remind the reader that these parameter values correspond
to the two exemplary cases denoted as (B) and (C) in Fig.~\ref{fig1}(A).
Several observations can be made by inspecting the figure and we
mention a few important ones.
Firstly, the inhomogeneous nature of the phase spaces
is reflected by the correlator. Specifically, the large
values of $C_{z}$ in the regular islands on the periphery of the
phase space imply participation of ``fast moving" Floquet
states in the dynamics. Examples of such Floquet states can be
seen in Fig.~\ref{fig4}(A) as states with large level velocities.
Secondly, the correlator takes on different values even in the
chaotic regions and therefore points to
the different nature of the dynamics {\it i.e.,} different groups
of Floquet states that influence the dynamics.
Finally, 
despite the similar looking phase spaces, it is clear from Fig.~\ref{fig5}
that the correlator maps are different. 
In particular, note that the correlator values in Fig.~\ref{fig5}A
about the left and the symmetry related right regular islands 
have the same sign. However, in Fig.~\ref{fig5}B the same regions have
correlators of opposite signs. This indicates a break in symmetry and
hence provides some hints to the fact that one obtains control
in this case as opposed to in case A. 

Before proceeding further
we note that the correlator is related
to the generic landscape functionals employed in OCT studies\cite{cr07}.
In order to see this consider the generic quantum control landscape
functional\cite{cr07}
\begin{equation}
\Phi(U) = {\rm Tr}\left[U(\tau) \rho(0) U(\tau)^{\dagger} \Theta \right]
\end{equation}
with $\rho(0), \Theta$, and $\tau$
being the initial density matrix, 
an arbitrary observable operator, and final dynamical time respectively.
Choosing $\rho \equiv |z\rangle \langle z|$ and 
$\Theta \equiv {\cal E}_{2\omega}(x,t) \equiv 
x \cos(2\omega t + \phi)$
one can express the landscape functional as
\begin{equation}
\Phi(U) = \sum_{n} \langle \chi_{n}(0)|z(\tau)\rangle
\langle z(\tau)|{\cal E}_{2\omega}|\chi_{n}(0)\rangle
\end{equation}
Using Eq.~\ref{zper} one can write
\begin{eqnarray}
\Phi(U) &=& \sum_{m,n} e^{-i(E_{n}-E_{m})\tau} 
\langle \chi_{n}|z\rangle \langle z|\chi_{m}\rangle 
\langle \chi_{m}|{\cal E}_{2\omega}(\tau)|\chi_{n}\rangle \nonumber \\
&=& \sum_{n} p_{zn} \left(\frac{\partial E_{n}({\bm \lambda})}{\partial
\lambda_{2}}\right) + {\rm nondiagonal\,\,terms}
\label{corrphirel}
\end{eqnarray}
The first term on the last line in Eq.~\ref{corrphirel} comes about
due to the Hellman-Feynman theorem. The diagonal term in
Eq.~\ref{corrphirel} is essentially the correlator in Eq.~\ref{intvelcor}
and the contribution from the nondiagonal terms
can be neglected in the long time limit. The variances in Eq.~\ref{intvelcor} 
ensure that $C_{z}$ is a true correlation function. However, without
the variances one is dealing with the covariance which is related to
the landscape functional $\Phi$ via Eq.~\ref{corrphirel}.
Note that the covariance already has the essential features that are
needed for our study and the existing semiclassical
insights\cite{kct02} should prove valuable in the context of the present work. 

\begin{figure*} [hbtp]
\includegraphics[height=70mm,width=180mm]{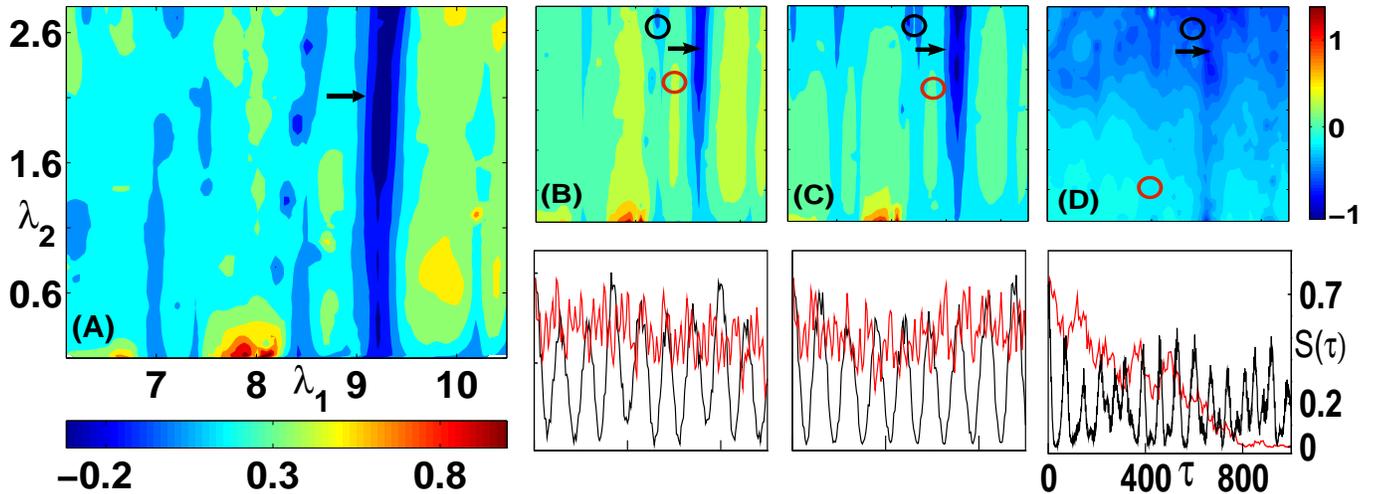}
\caption{(Color online) Control landscapes constructed via
the intensity-level velocity correlator
$C_{z}({\bm \lambda})$ for phase $\phi$ equal to (A) $0$ (B) $\pi/4$
(C) $\pi/3$ and (D) $\pi/2$. All landscapes shown have the same axes range.
Regions lacking control are 
associated with $C_{z}(\bm{\lambda}) \lesssim 0$. 
Note the one-to-one correspondence between the landscape in (A) and
the one shown in Fig.~\ref{fig3} and
the robustness of the
``wall'' of no control around $\lambda_{1} \approx 9.2$
(indicated by arrows). Survival probabilities shown in the lower panels
for specific field parameters (indicated by circles) confirm
the predictions based on $C_{z}(\bm{\lambda})$. Calculations of the
landscape are performed on a $50 \times 30$ grid with a spacing of $0.1$.}
\label{fig6}
\end{figure*}

What should one expect from the correlator? 
In general, $C_{z}(\bm{\lambda})$ highlights the presence of
groups of states exhibiting common localization features\cite{Toms96,kct02}.
In the ergodic limit it is known\cite{Toms96} that
$C_{z}(\bm{\lambda}) \sim 0 \pm N^{-1/2}$ for any initial state $|z\rangle$
with $N$ being the number of states involved in the averaging in
Eq.~\ref{intvelcor}. 
However, despite the strong fields and extensive chaos, the system
considered herein is not ergodic (cf. Fig.~\ref{fig5}).
The interpretation is subtle as 
compared to the earlier studies\cite{Toms96,kct02}
since chaotic states already exist for $\lambda_{2}=0$.
The chaotic states, due to
their delocalized nature, have small level-velocities
$|\partial E_{n}/\partial \lambda_{2}|$ which is evident from
Fig.~\ref{fig4}A as well. 
Consequently, participation of several chaotic
states in the decay dynamics of $|z\rangle$ results in small 
value for $|C_{z}(\bm{\lambda})|$. 
As a simple model,  assume $N_{c}$ chaotic states with nearly the same
level velocities $v_{c}$ contribute equally to the dynamics of $|z\rangle$
{\it i.e.,} similar intensities $p_{zc} = N_{c}^{-1}$. Note that the
division of the total number of states as $N = N_{c} + N_{r}$, with
$N_{r}$ being the number of regular states,
is strictly valid only in the semiclassical
limit. Clearly the parameters in this work are such that
the system is far from the required limit but the arguments are presented with
an attempt to understand some of the features of the correlator. Within
the simple model, using the sum rule that the average of the level-velocities
over all $N$ states is zero, it is possible to show that
\begin{eqnarray}
\langle v \rangle_{r} &\equiv& \frac{1}{N_{r}} \sum_{j=1}^{N_{r}} v_{j}
= -v_{c} f_{\rm ch} \\
C_{z} &=& v_{c}
\left[\frac{f_{\rm ch}(1+f_{\rm ch})}{v_{c}^{2}f_{\rm ch}+\langle v^{2}
\rangle_{r}}\right]^{1/2} 
\approx -\frac{\langle v \rangle_{r}}{\langle v^{2} \rangle_{r}^{1/2}}
\label{intvelcorappx}
\end{eqnarray}
with $f_{\rm ch} \equiv N_{c}/N_{r}$ and the limiting form of the correlator
above arises when $v_{c}^{-2}\langle v^{2} \rangle_{r} \gg f_{\rm ch} \gg 1$.
It is crucial to observe that
the sign of the correlator in Eq.~\ref{intvelcorappx}
is determined by the chaotic 
level-velocity $v_{c}$. An {\it a priori} knowledge of the sign of $v_{c}$
in a specific system is difficult but can be determined from
a single Floquet calculation.
Eq.~\ref{intvelcorappx} suggests
that $|C_{z}|$ will be small for $\langle v \rangle_{r} \ll
\langle v^{2} \rangle_{r}^{1/2}$.

\section{Interpretation of the features on the control landscape}
\label{results}

Armed with the qualitative insight into the behavior of
$C_{z}({\bm \lambda})$ we now compute the control landscape 
using Eq.~\ref{intvelcor} for 
the entire range of $\lambda_{1}$ shown in Fig.~\ref{fig1}A.
The resulting control landscape is shown in Fig.~\ref{fig6}A 
for relative phase $\phi=0$ and should be compared to the
landscape in Fig.~\ref{fig3} mapped via the time-smoothed
survival probability $\langle S \rangle$.
It is important to note that the two landscapes show identical
features. In particular, the regions lacking control on the
landscape in Fig.~\ref{fig3} correspond to the regions
showing $C_{z}(\bm{\lambda}) \lesssim 0$ in Fig.~\ref{fig6}A.
However, using the correlator based landscape it is now possible
to provide an explanation, not readily forthcoming from
the landscape in Fig.~\ref{fig3}, for the source of such lack of
control. 
For the specific parameters chosen in this study
the chaotic level-velocity distribution 
is peaked around a negative value and
thus $C_{z}(\bm{\lambda}) \lesssim 0$, following the 
results of the simple model in Eq.~\ref{intvelcorappx}, implies that
the regions of no control arise due to the participation of chaotic
Floquet states. Although the
role of chaotic states to control is already
highlighted in Fig.~\ref{fig4} at a single point on the
control landscape, it is worth mentioning that the
correlator identifies such regions on the landscape with remarkable ease.

A noteworthy feature on both the control landscapes 
is the existence of ``walls" of no control.
Interestingly, the wall
at $\lambda_{1} \approx 9.2$
is particularly robust and, hence,  
consistent with the results in Fig.~\ref{fig1} and Fig.~\ref{fig4}
which suggest extensive participation by delocalized Floquet states.
Therefore our observations establish $C_{z}({\bm \lambda})$
to be a highly sensitive measure capable of
identifying the lack of
control due to the involvement of chaotic states. 
Several other regions of no control can be seen in Fig.~\ref{fig6}A 
which are less robust in comparison
to the one at $\lambda_{1} = 9.2$. For instance, the
walls around $\lambda_{1} \approx 7.0$, and $8.2$
seem to break up at certain values
of $\lambda_{2}$ but, interestingly, reappear at higher $\lambda_{2}$. 
In every instance where $C_{z}({\bm \lambda}) \lesssim 0$ we have confirmed
the presence of atleast one chaotic state with substantial $p_{zn}$.
This points to a fairly nontrivial role played by the
chaotic states in the control process. The walls and islands of
no control result in the highly convoluted topology of the control
landscape.
Hence, using the symmetry breaking property of the
$2\omega$-field for control purposes is not very effective when
chaotic Floquet states are participating in the dynamics. 
On the other hand regions 
with $C_{z}(\bm{\lambda}) > 0$ in Fig.~\ref{fig6}A indicate varying
degree of control. Interestingly, an
earlier analysis\cite{fm93}, corresponding to case (C) in
Fig.~\ref{fig1}, happens to be in a region
away from the walls of no control! Again, the large
positive value of the correlator indicates
the absence of interactions with chaotic states leading to localization
due to symmetry breaking as seen in Fig.~\ref{fig4}C.

Is it possible to break the wall of no
control with a judicious choice of $\phi$? 
In order to explore such
a possibility we show the control landscape for different choices
of $\phi$ in Fig.~\ref{fig6}B, C, and D.
The case $\phi=\pi/2$ is
shown in Fig.~\ref{fig6}D and one observes that
$C_{z}(\bm{\lambda}) \lesssim 0$ over most of the landscape which
indicates very little control. This is consistent with the fact
that the combined driving field 
is symmetric about the time axis for $\phi=\pi/2$. 
The control landscape for $\phi=\pi/3$, and $\pi/4$ are also shown
in Fig.~\ref{fig6}C and B respectively. Note that decreasing $\phi$
from the symmetric value of $\pi/2$ gradually leads to regions of control.
Further confirmation comes from the
survival probability data shown in Fig.~\ref{fig6} for two points on
each of the control landscapes with $\phi \neq 0$.
It is intriguing to see that the wall of no control seen
in Fig.~\ref{fig6}A for $\lambda_{1} \approx 9.2$
is present for all the values of $\phi$
shown here. The reasons for such robustness is not clear at the moment.
Note that such regions in Fig.~\ref{fig6} for $\phi \neq 0$ are
characterized by large negative values of $C_{z}({\bm \lambda})$.
This seems to contradict the earlier works\cite{Toms96,kct02} which associated
a large value for $|C_{z}|$ with extensive localization in phase space.
However, as stated earlier, in the present case the interpretations
are subtle with the sign of the chaotic level-velocity playing a key role.
Our preliminary studies indicate that in such regions the assumption
$\langle v \rangle_{r} \ll \langle v^{2} \rangle_{r}^{1/2}$ is weakened
presumably due to an increased fraction of chaotic states.
Further work is needed to understand this aspect.
Nevertheless, the numerical results presented in Fig.~\ref{fig6}
indicate that regions of no control are signalled
by $C_{z} \leq 0$ whereas $C_{z} > 0$ implies control.

\section{Conclusions}
\label{summary}

In summary, the strong field control landscapes
in Fig.~\ref{fig6} show that nearly
perfect control is obtained as long as
there is no involvement of chaotic states in the dynamics.
However, involvement of the chaotic states
leads to lack of control and such regions on the control landscape
are interspersed among controllable regions in a complicated fashion
giving rise to a highly convoluted landscape topology.
Some of these regions are particularly robust under field amplitude
and relative phase variations (cf. Fig.~\ref{fig6}).  
Given that
convoluted landscapes are expected in OCT studies\cite{sr91,sr93} on
classically chaotic systems, it is reasonable to expect that chaotic states
can be associated with certain suboptimal features on quantum control 
landscapes. 
For systems with effective $\hbar \ll 1$
one expects\cite{tu94} several such 
plateau regions and the effectiveness of the control
would depend sensitively on the initial state and field parameters.

We emphasize that the association of uncontrollable regions
on the landscape with influence of the chaotic Floquet states is made
possible solely due to the highly sensitive nature of the
intensity-velocity correlator 
$C_{z}({\bm \lambda})$ (cf. Eq.~\ref{intvelcor}).
The sensitivity of $C_{z}({\bm \lambda})$ to the
underlying phase space structures is evident from
Fig.~\ref{fig5} and makes it an ideal candidate
to understand the origins of suboptimal
structures on the control landscape from a classical-quantum
correspondence perspective. Moreover,
$C_{z}({\bm \lambda})$ can be obtained with a few 
Floquet calculations or via the parametric equations of
motion\cite{ggbm96,mmr95} and 
hence applicable to multidimensional
driven systems. A further attractive feature, from the point of
multidimensional systems,
is that the correlator is capable of establishing the influence of
the chaotic states on the control landscape without the need for
visualizing/classifying the states in the underlying phase space. 

Although the present work deals explicitly with a tunneling
system, many of our observations are valid in other contexts as well.
In general, controlling the decay
(dissociation, ionization) of a state is expected to be complicated
if one or several chaotic states participate in the dynamics. The
complications will be reflected in the form of the optimal fields.
This does not contradict the recent work\cite{taka07} by Takami and Fujisaki
since they are concerned with controlling the transition between
two random states in a completely chaotic system. It is the mixed
regular-chaotic phase space scenario, a generic one for molecular systems,
which is expected to exhibit some of the features seen in the current
study. Finally, it would be interesting to see if 
the influence of chaotic states are indeed reflected in the topology of
the quantum control
level sets in an OCT calculation. In this regard the present work suggests
that the intensity-velocity correlator could prove to be a useful tool.
We are currently exploring the usefulness of the correlator
in other systems and
as a tool in formulating local control strategies.
   
\section*{Acknowledgments}

SK gratefully acknowledges useful 
discussions with Prof. Harshawardhan Wanare.
It is a pleasure to thank Dr. Raj Chakrabarti for sharing
his insights into control landscapes and for his crucial comments
on a preprint version of the manuscript which led to the
results in Eq.~\ref{corrphirel}.
Astha Sethi is funded by a Fellowship 
from the University Grants Commission, India.

\end{document}